\documentclass{article}    

\usepackage{graphicx}          
\usepackage{epsfig}
\usepackage{graphicx}
\usepackage{makeidx}
\usepackage{multicol}
\usepackage{verbatim}
\usepackage{subfigure}
\usepackage{mathrsfs}
\usepackage{snapshot}

\usepackage{amsmath}

\usepackage{crop}
\crop

\usepackage{amssymb}
\usepackage{graphicx}

\def\E{\mathbb{E}}

\def\ba{\begin{array}}
\def\ea{\end{array}}
\def\bi{\begin{itemize}}
\def\ei{\end{itemize}}

\newtheorem{defn}{Definition}[section]

\newtheorem{thm}{Theorem}[section]

\newtheorem{cond}{Condition}[section]

\begin{document}

\title{\LARGE \bf Empirical characteristic function identification of linear stochastic  systems with possibly unstable zeros}
\date{}
\author{L\'aszl\'o Gerencs\'er and M\'at\'e M\'anfay}

\maketitle

The purpose of this paper is to adapt the empirical characteristic
function (ECF) method to stable, but possibly not inverse stable
linear stochastic system driven by the increments of a
L\'evy-process. A remarkable property of the ECF method for i.i.d.
data is that, under an ideal setting, it gives an efficient
estimate of the unknown parameters of a given parametric family of
distributions. Variants of the  ECF method for special classes of
dependent data has been suggested in several papers using the
joint characteristic function of blocks of unprocessed data.
However, the latter may be unavailable for L\'evy-systems. We
introduce a new, computable score that is essentially a kind of
output error. The feasibility of the procedure is based on a
result of Devroye on the generation of r.v.-s with given c.f. Two
special cases  are considered in detail, and the asymptotic
covariance matrices of the estimators are given. The present work
extends our previous work on the ECF identification of stable and
inverse stable linear stochastic L\'evy-systems, see
\cite{ECC_own}.

\section{Introduction}

L\'evy processes have been widely used to model phenomena arising in natural sciences, economics, financial mathematics, queueing theory and telecommunication \cite{wind},\cite{novikov},\cite{CONT-TANKOV-FinModJunp}.
The geometric Brownian motion, which is considered the classical model for modeling the dynamics of financial instruments, was introduced by Louis Bacehelier
\cite{Bacehelier}.
Although empirical studies found that the model's assumptions do not correspond with reality,
this model is still the accepted core model.
Recently a new model has been used to model stock dynamics, called the geometric
L\'evy processes obtained by taking the exponential of a L\'evy process.

In this paper we present an identification method that is inspired by the so-called empirical characteristic function (ECF for short) method and the output error identification method. The ECF method, which can be interpreted as the Fourier transform of a maximum likelihood method, see \cite{FEUER+MC}, was first applied to estimate the unknown parameters of a characteristic function using i.i.d. samples. Carrasco and Florens showed in \cite{CARRASCO-EFFEMPIRCHAR} that the ECF method gives an efficient estimator for the problem of identifying the characteristics of a distribution that lacks a probability density function, but has a computable characteristic function (c.f.). The ECF method has been adapted to identify the noise characteristics of linear systems, but the identification of the system dynamics is typically out of the scope of papers. Among the few papers that estimate the system parameters in \cite{calder_davis} the parameters of an ARMA process are identified using M-estimators with a given distribution on the driving noise. In \cite{quasi_max} Schlemm and Stelzer estimates the system parameters and the covariance of the noise for L\'evy-driven continuous-time ARMA processes using quasi maximum likelihood method. In \cite{ECC_own} both the system parameters and the noise parameters are estimated and it is showed that a properly adapted ECF method estimates the system dynamics more effectively than standard methods such as prediction error and quasi maximum likelihood.

The purpose of this paper is to extend the results presented in \cite{ECC_own} to finite dimensional stochastic L\'evy systems with unstable zeros. Recall that both the PE method and the ML method as presented in \cite{gerencs_ML} assume that the system is non-minimum phase, ie. it has an exponentially stable inverse. The same assumption is used in \cite{ECC_own}. In fact, the identification of finite dimensional linear stochastic systems with unstable zeros is barely discussed in the literature. A remarkable feature of the ECF method is that is naturally applicable to the identification of finite dimensional stochastic systems if properly adapted. Our starting point is the ECF method for dependent data, as presented in the literature, using blocks of data, see \cite{knight}. This idea is then extended by defining a c.f. in terms of data passed through a possibly non-FIR filter. A novel challenge of this approach is that the exact c.f. cannot be computed explicitly (which is the key assumption for the ECF methods). However, it is found that an unbiased estimator for the exact c.f. can be obtained under the assumption that we can simulate or system with arbitrary feasible choice of the system parameters $\theta$ and noise parameter $\eta.$ The latter assumption is not unrealistic in view of the procedure presented in \cite{devroye}.

Thus we finally arrive at a procedure which can be viewed as a statistical output error method. The actual data are compared to simulated data, and the parameters of the latter are adjusted so as to ensure a good fit in a statistical sense. The resulting method can be analyzed along the lines of the classic ECF, or rather GMM method.

In retrospect, our method also extends the classic ECF method for i.i.d. data for situations when the c.f. is not available explicitly, but we do have an unbiased estimator of it in terms of a parameter-dependent random variable, say $\xi(\eta)$, which is computable via a mechanism of the form
$$
\xi(\eta)=F(\rho,\eta),
$$
where $F$ is a fixed, known function of $\rho$ and $\eta,$ and $\rho$ is a fixed random variable with known distribution. The data are generated via a true $\eta^*$ and the problem is to identify $\eta^*$. The above problem formulation is perfectly in line with the problem of system identification  with $\rho$ denoting the input noise and $\eta^*$ denoting the system parameters.

\section{L\'evy processes}

A L\'evy process $(Z_t)$ is a continuous-time stochastic
process that has stationary an independent increments. Thus, the behavior of a L\'evy process shows several similarities with that of the Wiener process, but the trajectories of L\'evy process
may be discontinuous. 
For an excellent introduction to the theory of L\'evy processes
see \cite{Sato}.

One of the simplest but not trivial example for a L\'evy process is the compound Poisson process. It is a Poisson process with random,
independent and identically distributed jumps. By extending the idea of the construction of compound Poisson processes we obtain a more general class L\'evy
processes, the so called pure-jump processes, formally given by
\begin{equation}
\label{eq:Levy_def}
Z_t = \int_0^t \int_{{\mathbf R}^1} x N(ds, dx),
\end{equation}
where $N(dt,dx)$ is a time-homogeneous, space-time Poisson point-
process, that counts the number of jumps of size $x$ at time $t$. A
simple and elegant introduction to Poisson point-processes in a
general state-space is given in \cite{kingman}. 
A basic technical tool in the theory of L\'evy processes is the L\'evy measure. For pure-jump processes their L\'evy measure is defined using the intensity of $N(dt,dx)$. That intensity is formally
defined by $\E [N(dt,dx)],$ with $\E $ denoting expectation. Due
to time homogeneity, $\E [N(dt,dx)]$ can be written as
$$\E [N(dt,dx)] = dt \cdot \nu(dx),$$
where $\nu(dx)$ is the so-called L\'evy-measure of process $Z_t.$
%
%
%
%

Now the above representation of a pure-jump L\'evy process given
in (\ref{eq:Levy_def}) is mathematically rigorous if
\begin{equation}
\label{eq:INTEGRABILIY}\int_{{\mathbf R}^1}  \min (|x|,1) \nu(dx)
< \infty. \end{equation} In the area of financial time
series sample paths with finite variations are obtained for most
indices, as supported by empirical evidence, see \cite{CGMY}.
In \cite{CGMY} such finite variation processes are obtained when modeling indices.
It also worth noting that
(\ref{eq:INTEGRABILIY}) implies that for all $t<\infty$
\begin{equation}
\E |Z_t| < \infty.
\end{equation}


Since a L\'evy process $Z_t$ has independent and identically distributed increments its characteristic function can be
written in the form
\[
\E \left[e^{iuZ_t}\right] = e^{t \psi(u)}.
\]
Here $\psi(u)$ is called the characteristic exponent of $Z_t$.

\section{Examples for L\'evy processes in modeling}

%
%
%

\label{sec:levy_fin}

The compound Poisson process is continuous-time stochastic process defined by its rate $\lambda$ and its jump size distribution $F$ via
$$
Z_t=\sum_{i=1}^{N_t}X_i,
$$
where $(N_t)$ is a Poisson process with rate $\lambda,$ and $X_i$-s are i.i.d. random variables with distribution $F.$ Such processes are widely used for modeling purposes in queueing theory, for example see \cite{tien}.

Geometric L\'evy processes have been widely used recently to model price processes.
Several L\'evy process have been proposed by different authors.
%
The $\alpha$-stable process was used to
price dynamics of wool by Mandelbrot in \cite{MANDELBROT}. The $\alpha$-stable process is defined via the L\'evy measure
\begin{equation}
\label{eq:stable}
\nu(dx) = C^- |x|^{-1-\alpha} {\mathbf 1}_{x < 0 } dx+ C^+ |x|^{-1-\alpha} {\mathbf 1}_{x
> 0 } dx,
\end{equation}
%
with $0 < \alpha < 2.$

Carr, Geman, Madan and Yor in \cite{CGMY} argues that the so-called CGMY process is able to model several important characteristics of price dynamics. The CGMY process is also called as tempered stable process because it
is obtained by setting $C^- = C^+$ in (\ref{eq:stable}), and then, separately for negative and positive $x$-s, multiplying the
L\'evy-density of the original symmetric stable process with a
decreasing exponential. Using
standard parametrization the L\'evy measure of the CGMY process is given by
\begin{equation}
\label{eq:cgmy}
\nu(dx) = {\frac {C e^{-G |x|}} {|x|^{1+Y}}}  {\mathbf 1}_{x < 0 } dx +
{\frac {C e^{-M x}} {|x|^{1+Y}}} {\mathbf 1}_{x
> 0 } dx,
\end{equation}
where $C,G,M >0$, and $0 < Y <2$. Intuitively, $C$ controls the
level of activity, $G$ and $M$ together control skewness.
Typically $G > M$ reflecting the fact that prices tend to increase
rather than decrease. $Y$ controls the density of small jumps,
i.e. the fine structure. For $Y<1$ the integrability condition (\ref{eq:INTEGRABILIY}) is satisfied, thus the corresponding L\'evy
process is of finite variation. The characteristic function of a
CGMY process $Z_t$ with parameters $C, G, M$ and $Y$ is given by
\begin{equation*}
\exp \{ t C \Gamma (-Y) \left((M-iu)^Y- M^Y + (G+iu)^Y- G^Y\right) \},
\end{equation*}
where $\Gamma$ denotes the gamma-function.

Variance Gamma process (VG for short), which has been proposed by
Madan, Carr and Chang in \cite{CARR-VG}, is formally obtained by setting $Y=0$ in (\ref{eq:cgmy}).
The VG-process is best implemented as
a time changed Brownian motion with drift, where the time
change is a gamma process, which is essentially the
continuous time extension of the inverse of a Poisson process.
Let
$W_t(\theta,\sigma)$ be a Brownian motion with drift, i.e.:
$$W_t(\theta,\sigma)=\theta t+\sigma W_t,$$
 where the process $(W_t)$ is a standard Brownian motion. Let $\gamma_t(\mu,\nu)$ be a gamma process with mean $\mu$ and variance $\nu,$ i.e. $\gamma_t(\mu,\nu)$ is a stochastic process with independent gamma distributed increments. Then the VG process with parameters $\sigma,\nu$ and $\theta$ is defined as
$$X_t(\sigma,\nu,\theta)=W_{\gamma_t(1,\nu)}(\theta,\sigma).$$
That is, the VG process is a time-changed Brownian motion. The VG process can also be interpreted as the difference of two gamma processes.

\section{ECF method for i.i.d. data}
\label{sec:iid}
In this section we briefly describe the ECF method for i.i.d.
samples with a finite number of moment conditions, see \cite{CARRASCO-EFFEMPIRCHAR}.
A remarkable property of the ECF method is that, under an ideal
setting, it gives an efficient estimate of the unknown parameters
of a given parametric family of distributions, see
\cite{CARRASCO-EFFEMPIRCHAR}. This observation is best justified,
heuristically, by the reasoning of A.~Feuerverger and
P.~McDunnogh, see \cite{FEUER+MC}, showing that  the score
functions of the ECF method for i.i.d. samples are obtained via
the Fourier transform of the left hand side of the log-likelihood
equation.

Assume that we are given an i.i.d. sequence of observations $(r_1,
r_2, \ldots r_N),$ such that their characteristic function is
known in closed form up to an unknown $q$-dimensional parameter
vector, say $\eta$, the true value of which is $\eta^*$. Let these
characteristic functions be denoted by $\varphi(u, \eta).$
%
%
The basic idea of the ECF method is to estimate $\eta^*$ by a
value of ${\hat \eta}$ such that $\varphi(u, {\hat \eta})$ best
matches the empirical characteristic function to be defined below.
For this purpose let us take a finite set of $u$-s, say $u_1, ...,
u_M$, with $M > q$, and for any of these $u$-s and any $n=1,...,N$
define a score, or equivalently, a generalized (normalized) moment
function:
\[
h_n(u, \eta) = e^{iu r_n}-\varphi(u, \eta).
\]
Clearly, $h_n(u, \eta)$ is a score-function in the sense that
setting $\eta = \eta^*$ we get
$$
\E \left[h_n(u, \eta^*)\right]=0
$$
for all $u.$ The above equations constitute an over-determined
system of non-linear algebraic equations with $q$ unknowns and $M
> q$ equations.

Now, let us take the sample average of the above scores and
define, for any fixed $u,$ the averaged fitting error as
\[
\overline{h}(u, \eta)=\frac 1N \sum_{n=1}^N h_n(u, \eta).
\]
Now letting $u$ vary over the finite set $u_1,...,u_M$ we define
the $M$-vector
$$
\overline{h}(\eta)= (\overline{h}(u_1,
\eta),...,\overline{h}(u_M, \eta))^T.$$ Defining the $M$-vectors
$$
g(\eta) = \E \left[\overline{h}(\eta)\right],
$$
and once again note that $g(\eta^*)=0,$ and thus $\eta^*$
satisfies the over-determined system of algebraic equations
\begin{equation}
g(\eta)=0.
\end{equation}
Since $g$ is not computable we may consider an alternative,
approximating equation
$$
\overline{h}(\eta)= 0,
$$
which will typically have no solution, since $M > q.$ Therefore we
seek a least-square solution by minimizing the weighted cumulative
error
\begin{equation}
\label{eqn:Q_FINITE} V_N(\eta)=||K^{-1/2}
\overline{h}(\eta)||^2,
\end{equation}
where $K$ is an appropriate, $M \times M$ weighing matrix to be
chosen below.

It can be shown that this method gives an asymptotically efficient estimator
of $\eta^*,$ moreover a precise characterization of the estimation
error can be given along the lines given in \cite{gerencs_arma}. At
this time we restrict ourselves to presenting a heuristics for
computing the asymptotic covariance matrix of $\hat{\eta}_N.$
First note that the minimization of the least squares error
$V_N(\eta)$ is (almost) equivalent to setting its gradient equal
to $0$, yielding the following $p$ equations:
$$
{\overline h}^*_{\eta} (\eta) K^{-1} {\overline h} (\eta) = 0,
$$
where $H^*$ denotes the complex conjugate of the matrix $H.$ Here
we followed the convention that the gradient w.r.t. $\eta$ is a
row-vector, and thus ${\overline h}_{\eta}$ is an $M \times q$
matrix, while ${\overline h}^*_{\eta}$ is $q \times M.$ The left
hand side of the above equation can be considered as a new set of
exactly $q$ scores.

A simple heuristics shows that the random effects in ${\overline
h}^*_{\eta}$ are negligible, and thus, defining the $M \times q$
sensitivity matrix
$$
G = g_{\eta} (\eta^*),
$$
an asymptotically equivalent problem is obtained by considering
the set of $q$ equations
$$
G^* K^{-1} {\overline h} (\eta) = 0.
$$
he l.h.s. can be considered as a new set of scores. Its
expectation is given by
$$
G^* K^{-1} g (\eta),
$$
and thus the Hessian of the asymptotic cost function, equal to $\E
\left[V_N(\eta)/N\right],$ is given by
$$
T= G^* K^{-1} G.
$$
In order to calculate the normalized covariance of the new set of
scores note that the normalized $M \times M$ covariance matrix of
${\overline h} (\eta^*)$ is given componentwise as
\begin{equation*}
\begin{split}
C_{k,l} =\E \left[ ~h^*_{n}(u_k, \eta^*) h_{n}(u_l,\eta^*)\right]= \\
\varphi(u_k - u_l,\eta^*) - \varphi(u_k ,\eta^*) ~\varphi(-
u_l,\eta^*).
\end{split}
\end{equation*}
Then the normalized covariance of the new set of scores is
$$
S = G^* K^{-1} C K^{-1} G .
$$

Now, following standard arguments, such as the Taylor-series
expansion of $ G^* K^{-1} {\overline h} (\eta)$ around $\hat
\eta,$ we get that the the asymptotic covariance matrix of the
estimator $\hat{\eta}_N$ is given by
$$
\Sigma_{\eta \eta} = T^{-1} S T^{-1}.
$$
Substituting the expressions for $T$ and $S$ obtained above we
get:
$$
\Sigma_{\eta \eta} = (G^* K^{-1} G)^{-1} ~G^* K^{-1} C K^{-1} G
~(G^* K^{-1} G)^{-1}.
$$
Simple linear algebra arguments yield that $\Sigma_{\eta \eta}$ is
minimized for $K=C$ and with this choice we get that
the \emph{asymptotic covariance matrix of the estimate} $\hat{\eta}_N$ obtained by the ECF method for i.i.d. data with a finite number of moment conditions is
$$\Sigma_{\eta \eta} =  (G^* C^{-1} G)^{-1}.$$

As indicated in the beginning of this section, the above
procedure, with the choice $K=C$, is efficient under ideal
conditions. More precisely, the ECF method presented above using
the full continuum of of $u$-s, $- \infty < u < + \infty,$ and a
suitably modification of the operator $K=C$ to ensure that its
inverse is a bounded linear operator, is as efficient as the ML
method, see \cite{CARRASCO-EFFEMPIRCHAR}.

\section{ECF method for filtered data}

In this section we extend the ECF method to dependent data
obtained by taking an i.i.d. sequence and passing it through a
stable finite dimensional linear system. A practically interesting
object of study is a linear stochastic system driven by a
L\'evy-process, or rather the increments of a  L\'evy-process. 
We write the system in the form
\begin{equation}
\label{eq:disc_levy2} \Delta y=A(\theta^*, q^{-1}) \Delta Z,
\end{equation}
%
%
where the time range is $- \infty < n < + \infty.$ Here $\Delta
Z_n$ denotes the increment of a zero mean L\'evy process $(Z_t)$
over an interval $[(n-1)h,nh),$ with $h>0$ being a fixed sampling
interval. $(Z_t)$ itself is defined for $ -\infty < t < + \infty$,
and it is tied to $0$ at time $t=0$, i.e. $Z_0=0$.
The condition
$$
\E \left[\Delta Z_n \right]=0
$$
significantly facilitates the analysis of the forthcoming ECF
estimations methods, in analogy with the analysis of the ML
method, see \cite{gerencs_ML}. Although in generally not
satisfied by the L\'evy processes presented in Section \ref{sec:iid}, it can
be enforced by preprocessing our data, as is customary in classic
time series analysis.

The L\'evy-measure of $Z$ will be denoted by
$\nu(dx)=\nu(dx,\eta^*),$ where $\eta^*$ denotes an unknown
parameter-vector with a known open range, say $D_{\eta} \subset
\mathbb{R}^q.$ The system dynamics depends on some unknown
parameter-vector $\theta^*,$ taking its values from some known
open set $D_{\theta} \subset \mathbb{R}^p.$
Let $D_{\theta}^*$ and $D_{\eta}^*$ be
compact domains such that $\theta^* \in D^*_{\theta} \subset
{\rm int ~} D_{\theta}$ and $\eta^* \in D^*_{\eta} \subset
{\rm int~ } D_{\eta}.$

\begin{cond}
\label{cond:stable} The operator $A(\theta, q^{-1})$ is a stable,
rational function of the backward-shift operator $q^{-1}$ for all
$\theta \in D_{\theta}.$ Moreover $A(\theta, q^{-1})$ is
three-times continuously differentiable w.r.t. $\theta$ for $\theta
\in D_{\theta}$.
\end{cond}

The smoothness of $A(\theta, q^{-1})$ w.r.t. $\theta$ should be
interpreted as follows: there exists a state-space realization of
$A(\theta, q^{-1})$ such that the state-matrices are three-times
continuously differentiable w.r.t $\theta$ for $\theta \in
D_{\theta}$.


Note that we did not assume the {\it inverse stability} of the
operator $A(\theta^*, q^{-1}),$ in contrast to standard
identification methods such as PE or ML. In particular, our method
is suitable for the identification of moving average (MA) systems
with unstable zeroes.

\begin{cond}\label{cond:MOMENTS_of_NU} We assume that for  all $q \geq 1$
  \begin{equation}
  \label{eq:moments}
  \int_{|x| \ge 1}|x|^q\nu(dx)<+\infty.
  \end{equation}
Moreover, it is assumed that the driving noise $(Z_t)$ is a zero
mean process: $$\E\left[ Z_t \right]=0.$$
\end{cond}
Note that the condition $\E\left[ Z_t \right]=0$ is a useful
technical assumption even in the case of ML identification, see
\cite{gerencs_ML}. In particular, it ensures that the estimators
of the system parameters and the noise parameters will be
asymptotically uncorrelated.

Now we are in the position to apply the ECF method for dependent
data, following the literature, in our special case. Consider the
parametric family of systems (or equivalently time series)
\begin{equation}\label{eq:par_dep_system}
\Delta y(\theta,\eta)=A(\theta)\Delta Z(\eta),
\end{equation}
with the time $n$ taking its values in $ - \infty < n < + \infty.$
Note that for $(\theta,\eta) = (\theta^*,\eta^*)$ we recover our
observed data in a statistical sense. The ECF method proposed in
the literature, see \cite{knight},\cite{FEUER_JASA},
%
%
is based on the computation of the joint characteristic function
of blocks of unprocessed data, i.e. for blocks of $(y_n).$ While
this computation can indeed can be carried out for special cases,
such as for Gaussian or stable noise processes, the computation of
the joint characteristic function is far from trivial in general.
One of the main contributions of this paper is to address this
challenge.

For a start, fix a block length, say $r,$ and define the
$r$-dimensional blocks
$$\Delta Y^r_n(\theta,\eta)=(\Delta y_{n-1}(\theta,\eta),\ldots,\Delta y_{n-r}(\theta,\eta)).$$
Then the joint characteristic function of the block $\Delta Y^r_n (\theta,\eta)$,
with $u=(u_1,...,u_r)^T$ being an arbitrary vector in $\mathbb
R^r,$
is given by
$$
\varphi_n(u,\theta,\eta)=\E \left[e^{iu^T \Delta
Y^r_n(\theta,\eta)}\right] = \E \left[e^{i \sum_{j=1}^r u_j \Delta
Y_{n-j}(\theta,\eta)}\right].
$$
Now, this can be explicitly computed, at least in theory. Letting
$h_l(\theta),~l=0,1,... $ denote the impulse responses of the
system $A(\theta),$ we can write
\begin{equation}
\begin{split}
\varphi_n(u,\theta,\eta) = \E \left[\exp \left\{ i\sum_{j=1}^r u_j \sum_{l=0}^{\infty} h_l(\theta) \Delta Z_{n-j-l}(\eta)\right\}\right]=  \label{eq:E_inf_prod}\\
\E \left[\exp \left\{ i \sum_{k=1}^{\infty}\Delta Z_{n-k} (\eta) \sum_{l \ge 0, j +l=k} u_j h_l(\theta) \right\}\right]. \\
\end{split}
\end{equation}
Fix $k$ and consider the last term. Setting $l = k-j$ introduce
the notation
$$
v_k(\theta) = \sum_{j=1}^{r} h_{k-j}(\theta) u_{j},
$$
with $h_l(\theta) =0$ for $l < 0.$ Then $v$ is the convolution of
$h$ and $u$:
$$
v=h * u.
$$
Denoting the characteristic function of $\Delta Z_n(\eta),$ for
any $n,$ by $\varphi_{\Delta Z(\eta)},$ we get
\begin{equation}
\label{eq:inf_prod}
\varphi_n(u,\theta,\eta) = \prod_{k=1}^{\infty} \varphi_{\Delta
Z(\eta)}(v_k(\theta)).
\end{equation}

Now the ECF method could be defined by fitting this theoretical
joint characteristic function to the empirical joint
characteristic function. Without providing details we point out
that it is not clear how to use such a procedure it in actual
computations, since $\varphi_n(u,\theta,\eta)$ is given in terms
of an infinite product. To circumvent this difficulty let us
return to the the definition of $\varphi_n(u,\theta,\eta)$. Note
that a simple unbiased estimation of $\varphi_n(u,\theta,\eta)$ is
given by
%
%
$$
e^{iu^T \Delta Y^r_n(\theta,\eta)} =  e^{i \sum_{j=1}^r u_j \Delta
Y_{n-j}(\theta,\eta)}.
$$
We propose to fit this theoretical vale to the data, and introduce
the scores
%
\begin{equation}
h_{n}(u, \theta,\eta)=e^{iu^{T} \Delta y_{n}}-e^{iu^{T} \Delta
y_{n}(\theta,\eta)}.
\end{equation}
Note that the score is essentially a kind of output error. Thus
the proposed procedure will be a generalization of the output
error identification method for the case when actual the input
process is not observed, but statistically known if $\eta^*$ is
known.

Note also that we can write the scores in the form
\begin{equation}
h_{n}(u, \theta,\eta)=e^{i ~(u * \Delta y)_{n}}-e^{i ~(u * \Delta
y(\theta,\eta))_{n}},
\end{equation}
where $u$ denotes the sequence $u_1,...,u_r$. The advantage of
this representation is that, in theory, we can use infinite
sequences of $u$-s representing the impulse responses of a finite
dimensional stable linear filter.

A final note: in order to compute the above score functions we one
have to be able to generate the i.i.d. noise sequence $\Delta
Z_n(\eta)$ for any given $\eta,$ having a prescribed c.f. $\phi(u,
\eta).$
This problem has been addressed and solved in \cite{devroye}.


To see the details of our procedure, suppose that we are given a
sequence of observed data $\Delta y_1,\ldots,\Delta y_{N+r}$ being
the outputs of (\ref{eq:par_dep_system}) with
$\theta=\theta^*,\eta=\eta^*.$ Construct the blocks of
observations $\Delta Y^r_n=(\Delta y_{n-1},\ldots,\Delta y_{n-r})$
for each $r < n \leq N+r.$ Take a set of vectors of dimension $r,$
say
$u_1,\ldots, u_M.$ Define the score functions as follows
\begin{equation}
h_{k,n}(\theta,\eta)=e^{iu_k^{T} \Delta y_{n}}-e^{iu_k^{T} \Delta
y_{n}(\theta,\eta)}
\end{equation}
for $k=1,...,M$ and $n=1,...,N.$ Note that these are indeed
appropriate score functions because
$$
\E \left[ h_{k,n}(\theta^*,\eta^*)\right]=0.
$$
The sample average of the scores is defined for any fixed $u_k$ as
\begin{equation}
\overline{h}_{k}(\theta,\eta)=\frac{1}{N} \sum_{n=r+1}^{N+r} h_{k,n}(\theta,\eta).
\end{equation}
Collecting the above sample averages over $k$ we define the
$M$-vector
\begin{equation}
\overline{h}(\theta,\eta)=(\overline{h}_{1}(\theta,\eta),...,\overline{h}_{M}(\theta,\eta))^T.
\end{equation}
Let $g(\theta,\eta)$ denote the expected error, i.e. let
\begin{equation}
g(\theta,\eta)=\E \left[ \overline{h}(\theta,\eta) \right].
\end{equation}
Clearly $\theta=\theta^*,\eta=\eta^*$ solves the over-determined
system of $M$ equations
$$
g(\theta,\eta)=0.
$$
Since $g$ is not computable we consider the alternative,
approximating equation
$$
\overline{h}(\theta, \eta)= 0,
$$
which will typically have no solution when $M > p + q.$ Therefore
we seek a least-square solution by minimizing the weighted
cumulative error
\begin{equation}
\label{eqn:Q_FINITE2} V_N(\theta, \eta)=||K^{-1/2}
\overline{h}(\theta, \eta)||^2,
\end{equation}
where $K$ is an appropriate, $M \times M$ weighing matrix to be
chosen below.

Instead of solving the minimization problem we define the
estimated parameter vectors $\hat{\theta}_N,\hat{\eta}_N$ as the
solutions of the gradient equation
\begin{align}
V_{\theta N}(\theta,\eta)&=0 \\
V_{\eta N}(\theta,\eta)&=0.
\end{align}

Instead we concentrate on the identification of the system dynamics. Suppose that the noise characteristics is given in such a form that it makes possible the generation of $Z(\eta).$ We construct the identification procedure along the just presented idea.

\section{Estimating the system dynamics}

Thus, suppose now that $\eta^*$ is known and we are able to
generate a sequence of i.i.d. random variables statistically
equivalent to $\Delta Z(\eta^*).$ With a slight abuse of notations
we shall use the same notations for real and simulated noise
sequences. Define the family of time-series parameterized by
$\theta$ as follows:
\begin{equation}\label{eq:par_dep_system2}
\Delta y_n(\theta)=A(\theta)\Delta Z_n(\eta^*),
\end{equation}
with $ - \infty < n < + \infty.$ Again for $\theta=
\theta^*$ we recover our observed data in a
statistical sense.
The score functions are defined as
\begin{equation}
h_{k,n}(\theta)=e^{iu_k^{T} \Delta y_{n}}-e^{iu_k^{T} \Delta y_{n}(\theta)}.
\end{equation}
One could easily mimic the steps of the construction of $V_{N}(\theta,\eta)$ to define $V_{N}(\theta).$ Again, fix a finite set of $u$-s, say $(u_1,\ldots,u_M).$ Define the average error for $u_k$
\begin{equation}
\overline{h}_{k}(\theta)=\frac{1}{N} \sum_{n=r+1}^{N+r} h_{k,n}(\theta).
\end{equation}
Let us define
\begin{equation}
\overline{h}(\theta)=\left(\overline{h}_{1}(\theta),\ldots,\overline{h}_{M}(\theta)\right)^T.
\end{equation}
$g(\theta)$ denotes the expected value of $\overline{h}$:
\begin{equation}
g(\theta)=\E \left[ \overline{h}(\theta) \right]
\end{equation}
Clearly $\theta=\theta^*$ solves the over-determined system of equations
$$
g(\theta)=0.
$$
By approximating $g$ by $\overline{h}$ we define $\hat{\theta}_N$ as the solution of
$$
V_{\theta N}(\theta)=||K^{-1/2}\overline{h}(\theta)||^2,
$$
where $V_N$ is the cost function defined by
$$
V_{N}(\theta)=||K^{-1/2}\overline{h}(\theta)||^2.
$$
The asymptotic score function is then defined as
$$
W(\theta)=\lim_{N \rightarrow \infty} \E \left[V_{N}(\theta)\right]=||K^{-1/2}g(\theta)||^2.
$$
\begin{cond}\label{cond:W_theta_solution}
$\theta^*$ is the unique solution of $W_{\theta}(\theta)$ in $D_{\theta}.$
\end{cond}
Following the arguments given \cite{gerencs_arma} we get the
following result:

\begin{thm}
Under Conditions \ref{cond:stable}, \ref{cond:MOMENTS_of_NU} and \ref{cond:W_theta_solution} we have
$$
\hat{\theta}_N-\theta^*=W_{\theta \theta}^{-1}(\theta^*)V_{\theta
N}(\theta^*)+O_M(N^{-1}).
$$
\end{thm}


Now we are ready to calculate the asymptotic covariance of the
estimator. Let $\Lambda'$ be the $M \times M$ covariance matrix with
entries
$$
\Lambda'_{k,l} = \E \left[ ~h^*_{k,n}(\theta^*) h_{l,n}(\theta^*)
\right].
$$

\begin{thm}
Under Conditions \ref{cond:stable}, \ref{cond:MOMENTS_of_NU} and
\ref{cond:W_theta_solution} the asymptotic covariance matrix of
$\hat{\theta}_N$ with the optimal choice of the weighting matrix $K=\Lambda'$ is
given by
$$
\Sigma_{\theta \theta} = 2 (H^* \Lambda^{-1} H)^{-1},
$$
where the $k^{th}$ row of $H$ is given by
$-\varphi_{\theta}(u_k,\theta^*,\eta^*),$ and $\Lambda$ is an $M \times M$ matrix with entries
$$
\Lambda_{k,l} = \varphi(u_k - u_l,\theta^*,\eta^*) - \varphi(u_k ,\theta^*,\eta^*) \varphi(- u_l,\theta^*,\eta^*).
$$

\end{thm}

{\bf Proof:}

The asymptotic gradient is given by
$$
g^*_{\theta}(\theta^*) K^{-1} g (\theta^*),
$$
while its derivative w.r.t. $\theta$ at $\theta^*$ (the Hessian of the asymptotic cost) is
$$
R^*= g^*_{\theta} (\theta^*) K^{-1} g_{\theta} (\theta^*).
$$
Then the Hessian of the asymptotic cost is
$$
T= H^* K^{-1} H.
$$
Note that since $\Delta y_n$ and $\Delta y_n(\theta^*)$ are independent as they are generated using different $\Delta Z_n$ sequences we have
$$
\Lambda'_{k,l} = 2\left(\varphi(u_k - u_l,\theta^*,\eta^*) - \varphi(u_k ,\theta^*,\eta^*) \varphi(- u_l,\theta^*,\eta^*)\right).
$$
We note in passing that $\Lambda'=2\Lambda.$ Thus the asymptotic covariance of the new set of scores is
$$
S = H^* K^{-1} \Lambda' K^{-1} H .
$$
The asymptotic covariance of the estimator $\hat \theta_N$ is then
$$
(H^* K^{-1} H)^{-1} ~H^* K^{-1} \Lambda' K^{-1} H ~(H^* K^{-1} H)^{-1}.
$$
It is easy to see that the optimal value of $K$ is
$$
K= \Lambda'
$$
yielding the asymptotic covariance for $\hat \theta_N$
$$
\Sigma_{\theta \theta}=(H^* \Lambda'^{-1} H)^{-1}=2(H^* \Lambda^{-1} H)^{-1}.
$$

Recall that $H=g^*_{\theta} (\theta^*),$ so that the $k^{th}$ row of $H$ is
$$
\frac{\partial}{\partial \theta} \E \left[ \overline{h}_k(\theta) \right] |_{\theta=\theta^*}=- \varphi_{\theta}(u_k,\theta^*,\eta^*).
$$
Hence, using the full continuum of moment conditions would yield the asymptotic covariance presented in \cite{CARRASCO-EFFEMPIRCHAR}, which implies the identification method in question is efficient.

\emph{Remark:}
The covariance matrices $\Sigma_{\theta \theta}=2(H^* \Lambda^{-1} H)^{-1}$ and $\Sigma_{\eta \eta}=(G^* C^{-1} G)^{-1}$ have similar structure. The rows of $H$ and $G$ are derivatives of the characteristic function of the observed data with respect to the unknown parameters $\theta$ and $\eta,$ respectively. Both $\Lambda$ and $C$ have entries of the form $$\varphi(u_k-u_l)-\varphi(u_k)\varphi(-u_l),$$ here $\varphi$ denotes the characteristic function of the observed data.

\section{ECF for i.i.d. data revisited}

In this section we give an extension of the ECF method for i.i.d. data under the assumption that the c.f. is not known explicitly, but we do have a computable random variable $\xi(\eta)$ such that
$$
\varphi(u,\eta)=\E\left[ e^{iu\xi(\eta)}\right].
$$
More exactly, we assume that we have a mechanism to compute an i.i.d. sequence $\xi_n(\eta)$ given by
$$
\xi_n(\eta)=F(\rho_n,\eta),
$$
where $\rho_n$ is an i.i.d. sequence that we can generate, and $F$ is a known function of $\rho$ and $\eta$, which is sufficiently smooth in $\eta$.

Let the true parameter be denoted by $\eta^*$, and let the observed sequence be
$$
\xi_n^*=F(\rho_n^*,\eta^*),
$$
where $(\rho_n^*)$ is a realization of an i.i.d. sequence with given distribution. The problem is then to identify $\eta^*$. The purpose of this exercise is to understand the problem if identifying the noise characteristic of a finite dimensional L\'evy system under a simpler settings. An obvious candidate for a score function is now
$$
h_n(u,\eta)=e^{iu\xi_n(\eta^*)}-e^{iu\xi(\eta)},
$$
where $\xi_n(\eta^*)$ are real data and $\xi_n(\eta)$ are simulated data. Taking a finite set $u$-s, say $u_1,\ldots,u_M,$ define
\begin{equation}
h_{k,n}(\eta)=e^{iu_k\xi_n(\eta^*)}-e^{iu_k\xi(\eta)}.
\end{equation}
From here we may proceed like in Section \ref{sec:iid} to define the quadratic cost function $V_{N}(\eta)$ and the corresponding objects $\overline{h}(\eta),$ its expected value $g(\eta)$ and $G=g_{\eta}(\eta^*).$ One could follow the line of reasoning presented in Section \ref{sec:iid} and obtain that the asymptotic covariance for the estimated parameter $\hat{\eta}_N$ is
\begin{equation}
\Sigma'_{\eta \eta}=2(G^* C^{-1} G)^{-1},
\end{equation}
where $C_{k,l}$ is defined in Section \ref{sec:iid} and the $k^{th}$ row of $G$ is
\begin{equation*}
\begin{split}
&-\left.\E\left[ \frac{\partial}{\partial \eta} e^{iu_k\xi(\eta)}\right]\right|_{\eta=\eta^*}=
-\left.\frac{\partial}{\partial \eta}\E\left[ e^{iu_k\xi(\eta)}\right]\right|_{\eta=\eta^*}=\\
&-\varphi_{\eta}(u_k,\eta^*).
\end{split}
\end{equation*}
For, computing the covariance of the scores gives
\begin{equation*}
\begin{split}
&\E \left[ ~h^*_{k,n}(\eta^*) h_{l,n}(\eta^*)\right]= \\
&2(\varphi(u_k - u_l,\eta^*) - \varphi(u_k ,\eta^*) \varphi(- u_l,\eta^*))=2C_{k,l}.
\end{split}
\end{equation*}
Comparing the variance of the ECF estimators for i.i.d. data yields the following result:
\begin{thm}
Denote the variance of the ECF estimator for i.i.d. data with known characteristic function presented in Section \ref{sec:iid} by $\Sigma_{\eta \eta}$ and denote the variance of the estimator for i.i.d. data without known characteristic function (but with a computable random variable) by $\Sigma'_{\eta \eta}$. Then we have
$$
2\Sigma_{\eta \eta}=\Sigma'_{\eta \eta}.
$$

\end{thm}
This result shows the change in the variance of the estimates caused by the fact that the c.f. is unknown.

\appendix


Let $\theta$ be a $d$-dimensional parameter vector.
\begin{defn}
We say that $x_n(\theta)$ is $M$-bounded if for all $q \geq 1$,
$$ M_q(x)=\sup_{n>0, \theta \in D} \E^{1/q}\left|x_n(\theta)\right|^q < \infty $$
\end{defn}
Define $\mathscr{F}_n=\sigma\left\{e_i: i \leq n \right\}$ and $\mathscr{F}^+_n=\sigma\left\{e_i: i > n \right\}$ where $e_i$-s are i.i.d. random variables.

\begin{defn}
We say that a stochastic process $\left(x_n(\theta)\right)$ is $L$-mixing with respect to $\left(\mathscr{F}_n,\mathscr{F}^+_n\right)$ uniformly in $\theta$ if it is $\mathscr{F}_n$ progressively measurable, M-bounded with any positive $r$ and
\begin{equation*}
\gamma_q(r,x)=\sup_{n \geq r, \theta \in D} \E^{1/q} \left|x_n(\theta)-\E\left[x_n(\theta)|\mathscr{F}^+_{n-r}\right]\right|^q,
\end{equation*}
we have for any $q \geq 1,$
$$\Gamma_q(x)=\sum_{r=1}^{\infty} \gamma_q(r,x) < \infty.$$
\end{defn}

Define $$\Delta x/\Delta^{\alpha} \theta=\left|x_n(\theta+h)-x_n(\theta)\right|/\left|h\right|$$
for $n\geq0, \theta\neq\theta+h \in D.$

\begin{thm}\label{thm:sup}
Let $(u_n(\theta))$ be an $L$-mixing uniformly in $\theta \in D$ such that $\E u_n(\theta)=0$ for all $n \geq 0, \theta \in D,$ and assume that $\Delta u /\Delta \theta$ is also $L$-mixing uniformly in $\theta, \theta+h \in D.$ Then
\begin{equation}
\sup_{\theta \in D_0} \left|\frac{1}{N} \sum_{n=1}^N u_n(\theta)\right|=O_M(N^{-1/2})
\end{equation}
\end{thm}

\end{document}